\documentclass[preprint2]{aastex}

\shorttitle{The Slitless Spectroscopy Software aXe}
\shortauthors{K\"ummel, M.\ et al.}

\begin{document}

\title{The Slitless Spectroscopy Data Extraction Software aXe}

\author{M.\ K\"ummel\altaffilmark{1}, J.R.\ Walsh\altaffilmark{1}, N.\ Pirzkal\altaffilmark{2}, H. Kuntschner\altaffilmark{1} \& A. Pasquali\altaffilmark{3}}
\altaffiltext{1}{Space Telescope -- European Coordinating Facility,
Karl-Schwarzschild-Str. 2, D-85748 Garching, Germany; mkuemmel, jwalsh, hkuntsch @eso.org}
\altaffiltext{2}{Space Telescope Science Institute, Baltimore, MD 21218, USA; npirzkal@stsci.edu}
\altaffiltext{3}{Max-Planck-Institut f\"ur Astronomie, K\"onigstuhl 17, D-69117 Heidelberg, Germany; pasquali@mpia-hd.mpg.de}

\begin{abstract}
The methods and techniques for the slitless spectroscopy software aXe,
which was designed to reduce data from the various slitless spectroscopy 
modes of Hubble Space Telescope instruments, are described. aXe can treat 
slitless spectra from different instruments such as ACS, NICMOS
and WFC3 through the use of a configuration 
file which contains all the instrument dependent parameters.
The basis of the spectral extraction within aXe are the position, morphology
and photometry of the objects on a companion direct image.
Several aspects of slitless spectroscopy, such as the overlap of 
spectra, an extraction dependent on object shape and the provision of 
flat-field cubes, motivate a dedicated software package, and the 
solutions offered within aXe are discussed in detail. The effect of the 
mutual contamination of spectra can be quantitatively assessed in aXe, 
using spectral and morphological information from the companion direct 
image(s). A new method named 'aXedrizzle' for 2D rebinning and co-adding spectral
data, taken  with small shifts or dithers, is described.
The extraction of slitless spectra 
with optimal weighting is outlined and the correction of spectra for 
detector fringing for the ACS CCD's is presented. Auxiliary software 
for simulating slitless data and for visualizing the results of an aXe 
extraction is outlined.
\end{abstract}




\section{Introduction} 
A spectrum produced by a dispersing element, usually a prism or a
grism, without a slit or slitlets has a number of characteristics that
require specialized data reduction tools. Slitless spectroscopy is
primarily a survey tool, since the spectra of all objects within a
given area defined by the telescope-instrument-detector combination
are recorded. While this survey aspect brings advantages, it also has
some drawbacks.
There can be no prior selection of target objects, and
contamination, which is an overlapping of spectra, occurs frequently.
The lack of defining slits
implies that for extended objects the spectral resolution is controlled by
the object size, more specifically the component of the object size in the
dispersion direction. Moreover the shape of the objects, which in general are
not aligned with the dispersion direction, must be taken into account during
the spectral extraction. Since in slitless spectroscopy each pixel can
receive radiation of any wavelength (within the sensitive range of the instrument),
special techniques for flat-fielding and fringing are required.
\begin{deluxetable}{llccc}
\tablecaption{The HST slitless spectroscopic modes to which aXe has been applied.\label{spec_modes}}
\tablehead{\colhead{Instrument/Camera}&\colhead{Disperser} & \colhead{Wav. Range [\AA]} & 
\colhead{Dispersion [\AA/pixel]} & \colhead{FOV\tablenotemark{1} [arcsecond]}}
\startdata
ACS/WFC     & G800L     & $5500-10500$     &     $38.5$   & $202\times 202$\\
ACS/HRC     & G800L     & $5500-10500$     &     $23.5$   & $29\times 26$\\
ACS/HRC     & PR200L     & $1700-3900$     & 20[@2500\AA]\tablenotemark{2} & $29\times 26$\\
ACS/SBC     & PR130L     & $1250-1800$     & 7[@1500\AA]\tablenotemark{2} & $35\times 31$\\
ACS/HRC     & PR110L     & $1150-1800$     & 10[@1500\AA]\tablenotemark{2} & $35\times 31$\\
NICMOS/NIC3 & G141      & $11000-19000$    & 80.0         & $51\times 51$          \\
WFC3/UVIS\tablenotemark{3} & G280      & $2000-4000$      & 13.0         & $160\times160$\\
WFC3/IR\tablenotemark{3} & G102      & $7800-10700$     & 25.0         & $123\times137$\\
WFC3/IR\tablenotemark{3}& G141      & $10500-17000$    & 47.0         & $123\times137$\\
\enddata
\tablenotetext{1}{Field-of-View}
\tablenotetext{2}{On account of the dispersion change with wavelength for prism 
spectra, the resolution
is accompanied by the wavelength it refers to (see also Eqn.\ \ref{eq_prdisp})}
\tablenotetext{3}{Prior to the WFC3 installation in Servicing Mission 4, aXe was used in the reduction of ground based
calibration data.}
\end{deluxetable}

Slitless spectroscopy has been employed for surveys from the ground
such as for the Hamburg/ESO objective-prism survey survey \citep{wi96},
the UK Schmidt Telescope (Clowes et al.\ 1980; Hazard et al.\ 1986) and
the Quasars near Quasars survey \citep{wo08}. Early efforts to use this
technique for the determination of stellar radial velocities are
described in \cite{fe69}. However, the strong disadvantage of
slitless spectroscopy from the ground is the high background, since the
lack of a slit implies that each pixel receives the full transmission
of the dispersing element. The background can be reduced by choosing a
high resolving power disperser or adding a narrow band filter to the
light path, such as for emission line studies on known lines (e.g.
Salzer et al.\ 2000; Jangren et al.\ 2005).

Deep slitless surveys from the ground, such as searches for high redshift
galaxies at $3 < z < 7$, are still valuable (see Kurk et al.\ 2004), however space based
observations are particularly effective, since the background is many orders of magnitude
lower than from  earth and there are no strong and variable
atmospheric absorption and emission components.
For the Hubble Space
Telescope (HST), the high spatial resolution (through lack of
atmospheric seeing) brings a further gain, since the background
contribution to the extracted spectrum is reduced. Although the
background in slitless mode is larger by the ratio of the spectrum
length to the slit width with respect to slit spectroscopy, HST
slitless spectroscopy can challenge the faint limits of ground-based
telescopes \citep[e.g. ][]{ma05}.

Surveys with HST in slitless mode have included the NICMOS/Hubble Space
Telescope Grism Parallel Survey of H$\alpha$ emission line galaxies in
the redshift range 0.7 to 1.9 with the G141 grism \citep{mc99}; the
Space Telescope Imaging Spectrograph Parallel Survey (Gardner et al.\
1998; Teplitz et al.\ 2003) of [O~II] and [O~III] emission line
galaxies $0.43<z<1.7$; and the Advanced Camera for Surveys GRAPES study
of the Ultra-Deep Field \citep{pir04} with, for example, detection of
Ly-$\alpha$ and Lyman-break galaxies to $z\sim6$ \citep{ma05}; the
PEARS survey in the GOODS North and GOODS South fields \citep{ma08} and
searches for high-z supernovae to $z\sim1.6$ (Riess et al.\ 2004;
Strolger et al.\ 2004).

After exploratory attempts to use standard spectral extraction tasks in
public data reduction packages (e.g., IRAF, IDL and MIDAS) for slitless
spectroscopic data, it was decided to develop a dedicated extraction
package.These standard packages encapsulate an enormous amount of expertise
for slit spectroscopy. However the lack of solutions for specific problems
in slitless spectroscopy together with the demand for the processing of
hundreds to thousands of spectra per image mandated a dedicated software
package. 

This software package, called aXe, was originally developed for
Advanced Camera for Surveys (ACS) grism and prism spectroscopy (see
Table \ref{spec_modes} and Pasquali et al.\ 2006). However by design
aXe was intended to be instrument-independent, and this aim was
achieved by placing all the instrument specific parameters into a
single data file called a {\it configuration file}. In addition to ACS,
aXe has subsequently been successfully applied to NICMOS
slitless data \citep{fr08} and data obtained during the ground calibration campaigns
of Wide Field Camera 3 (WFC3, Bond et al.\ 2007, Kuntschner et al.
2008b; see also Table \ref{spec_modes}).

This paper is devoted to a description of the aXe spectral extraction
package. aXe does {\it not}\/ work like a data reduction pipeline,
which has a static set of products as output. Depending on the data and
the scientific questions to be answered, the user can reduce the data
following different recipes and methods, and the subsequent Sections
illustrate all the methods and techniques implemented in the current
version aXe-1.6. Section
\ref{sect:axe_strat} is devoted to the extraction of spectra from an
individual slitless image. Section \ref{sect:contamination} explains the
various techniques to treat the contamination problem, followed by the
introduction of a 2D-resampling technique which aims at co-adding
individual spectra of the same object in Section \ref{sect:axedrizzle}.
The aXe implementation of optimal weighting is illustrated in Section
\ref{optimal_weighting}, followed by a discussion of the effects of
detector fringing (Section \ref{sect:fringing}). The structure of aXe
and related software packages for simulation and visualization are
presented in Section \ref{sect_axe_software_package}. The paper closes
with a summary and options for further refinement of the package.

\section{Extracting spectra from single slitless images}
\label{sect:axe_strat}
This section  details the aXe method to extract spectra from individual
slitless images. aXe requires associated direct images for the slitless images
to enable data processing.

\subsection{Slitless image -- direct image relation}
\label{sect:slitlessimage}
In slitless spectroscopy the location of spectra on the detector is
defined by the position of the object itself within the Field-of-View (FOV).
Therefore, in order to recover the absolute wavelength scale,
the position of the object
must be known in advance in the reference frame of the slitless
image. Although not a fundamental limitation, the presence of a
direct image from which the position of the object can be determined is
usually taken as a pre-requisite for the reduction of slitless spectra.
In instrumental configurations with a grism as disperser, the 0$^{th}$ order
spectrum, when available, could be
used to provide the zero point of the wavelength scale. However, due to
the prism on which the grating is ruled, the 0$^{th}$ order appears in
practice slightly dispersed which does not favour an accurate position
determination.

\begin{figure*}[t]
\includegraphics[angle=0, width=0.95\textwidth]{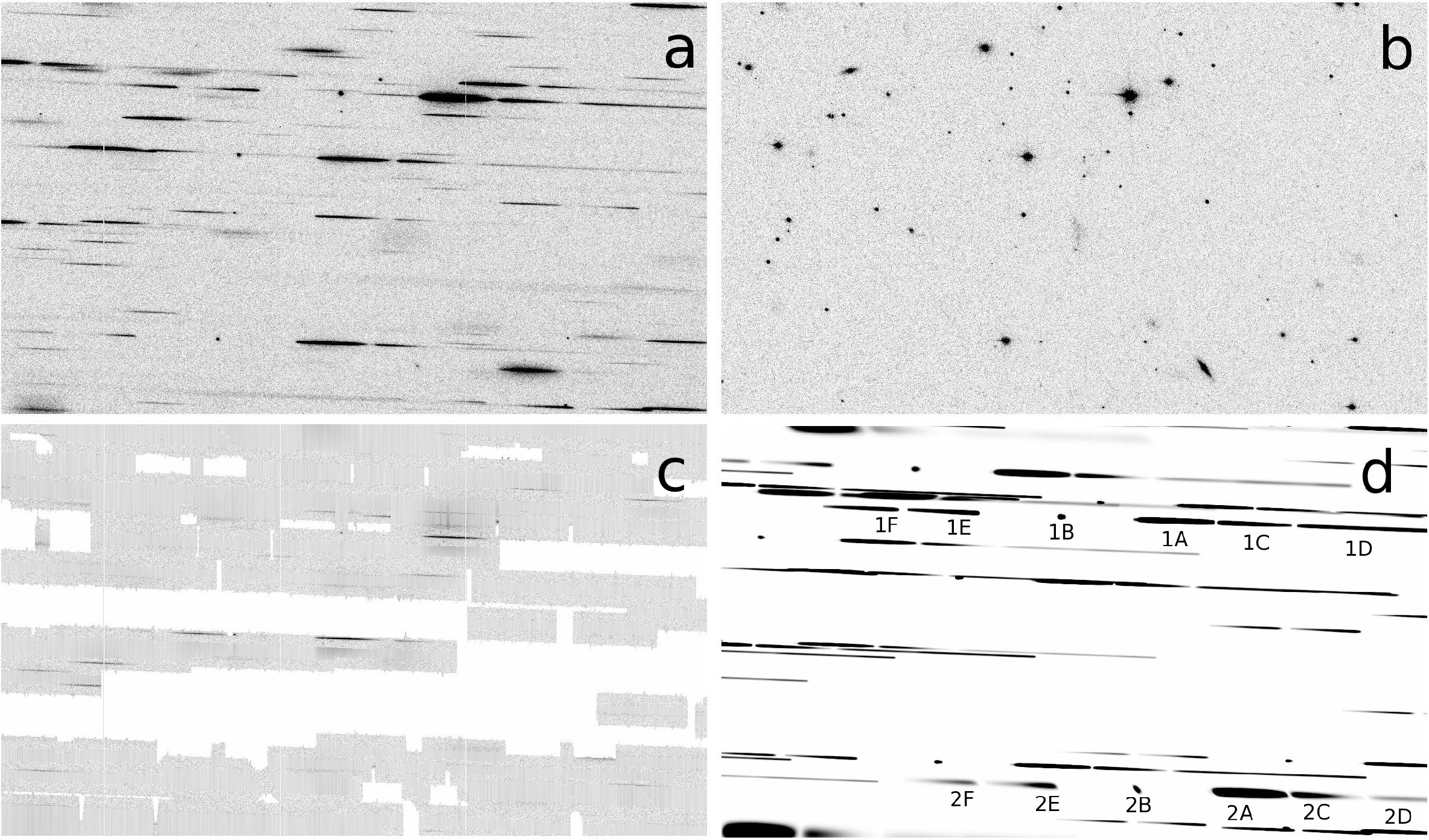}
\caption{Cutouts ($1200\times 700$ pixel) of an ACS/WFC G800L slitless image and the associated F850LP
direct image are shown on panels a and b, respectively to illustrate the input data for aXe.
Bad pixels and cosmic ray hits have been removed.
Panels c and d show intermediate results from a spectral extraction with aXe. The local background image
(see Sect.\ \ref{sect:loc_backgr}) is shown in panel c and a model of the slitless image, which is used to
estimate the contamination (see Sec.\ \ref{sect:quantcont}) in panel d. Since the extraction which produced
images c and d was limited to bright objects ($m_{AB} < 22.0$), some fainter sources are not masked out
in the background image and are not part of the model image in panel c. The $+1^{st}$ (A), $0^{th}$ (B),
$+2^{nd}$ (C), $+3^{rd}$ (D), $-1^{st}$ (E) and $-2^{nd}$ (F) order spectra for two objects
$1$ and $2$ are marked in the model image (panel d).
\label{fig:composit_image}}
\end{figure*}

In order to establish for each object a reference position on the slitless image, aXe requires, for
each slitless image (Figure \ref{fig:composit_image}, panel a), a corresponding direct
image of the same sky area (Figure \ref{fig:composit_image}, panel b). The pointing information
as given in the world coordinate system information in the headers of the slitless and direct images
allows to transform object positions from the direct images to the slitless images.
The reference positions are then used as the origins of local coordinate systems to describe,
for each object, its spectral properties.

\subsection{The aXe configuration and calibration files}
\label{sect:axeconfig}
aXe has been applied to the various HST slitless spectroscopic modes listed in
Table \ref{spec_modes}, which all have different properties,
e.g. different descriptions of the trace (the location of the center of gravity in the spatial direction)
and dispersion. In order to be able to reduce these distinct
instrumental modes with aXe, all information for a specific mode  is stored in a single
{\it configuration file}. An aXe configuration file contains:
\begin{itemize}
\item the data description (pointers to the science, error and data quality FITS image extensions);
\item the location and extent of the different spectral order(s);
\item the trace description for the spectral order(s);
\item the dispersion solution for the spectral order(s);
\item the names of the calibration files (flat-field and sensitivity files) to be used.
\end{itemize}

The trace description in aXe is given as a polynomial:
\begin{eqnarray}
\Delta y(\Delta x ) = a_0 + a_1 * \Delta x + a_2 * \Delta x^2 \ldots
\label{eq_trdesc}
\end{eqnarray}
with $(\Delta x, \Delta y) = (x-x_{ref}, y-y_{ref})$, the relative distance
of the image coordinates $(x, y)$ from the reference position of an object
on the slitless image $(x_{ref}, y_{ref})$.
Grism dispersion solutions are described as polynomials:
\begin{eqnarray}
\lambda(l) = l_0 + l_1 * l + l_2 * l^2 \ldots
\label{eq_grdisp}
\end{eqnarray}
with the distance along the trace $l$. Since prisms display quite rapid variation of
the spectral dispersion with wavelength, the functional
form for the dispersion solution is better described as:
 \begin{eqnarray}
\lambda(l) = l_1 + \frac{l_2}{(l-l_0)} + \frac{l_3}{(l-l_0)^2} + \frac{l_4}{(l-l_0)^3} \ldots
\label{eq_prdisp}
\end{eqnarray}
Around the singularity at $l=l_0$ the resolution decreases rapidly, causing a
pile-up in instrumental configurations such as ACS/HRC/PR200L which still have significant
sensitivity at the start wavelength of the pile-up \citep{la06}.

All quantities $a_0, a_1, a_2, \ldots, l_0, l_1, l_2, \ldots$ in Eqns.\ \ref{eq_trdesc}-\ref{eq_prdisp}
can be given as a 2D field dependent polynomial in order to take variations of the
trace description and dispersion solution as a function of object position
into account. This 2D field dependence is caused by field effects and the
different geometrical distortion between the direct and the dispersed images.
The quantity $a$ given as a 2D polynomial of order $2$ is, for example:
\begin{eqnarray}
a &=& \alpha_0 +                   \\ \nonumber
  & & \alpha_1 * x + \alpha_2 * y +\\ \nonumber
  & & \alpha_3 * x^2 + \alpha_4 * x*y + \alpha_5 * y^2
\label{eq_fdepend}
\end{eqnarray}
where $x$ and $y$ are the reference object position on the slitless image.

A thorough description of the aXe configuration files and the various quantities
declared therein can be found in the aXe Manual \citep{kue08}.
As calibration files aXe needs sensitivity files to transform the extracted
spectra from detector units ($electrons/s$) to flux units ($f_\lambda$) and
special flat-fields, which are introduced and described in Sect.\ \ref{sect:flatfield}.
The HST instrument science teams conduct dedicated observing programmes to
derive calibration files, trace description and dispersion solution
in the various slitless modes (see e.g.\ Pasquali et al.\ 2006; Kuntschner,
K\"ummel \& Walsh 2008a). aXe configuration and calibration files for all supported
instruments are available for download on the ST-ECF
webpages (\anchor{http://www.stecf.org/instruments/}{http://www.stecf.org/instruments/}).

\subsection{Spectral extraction}
\label{sect:spec_extr}
\begin{figure}[t]
\includegraphics[angle=0, width=0.45\textwidth]{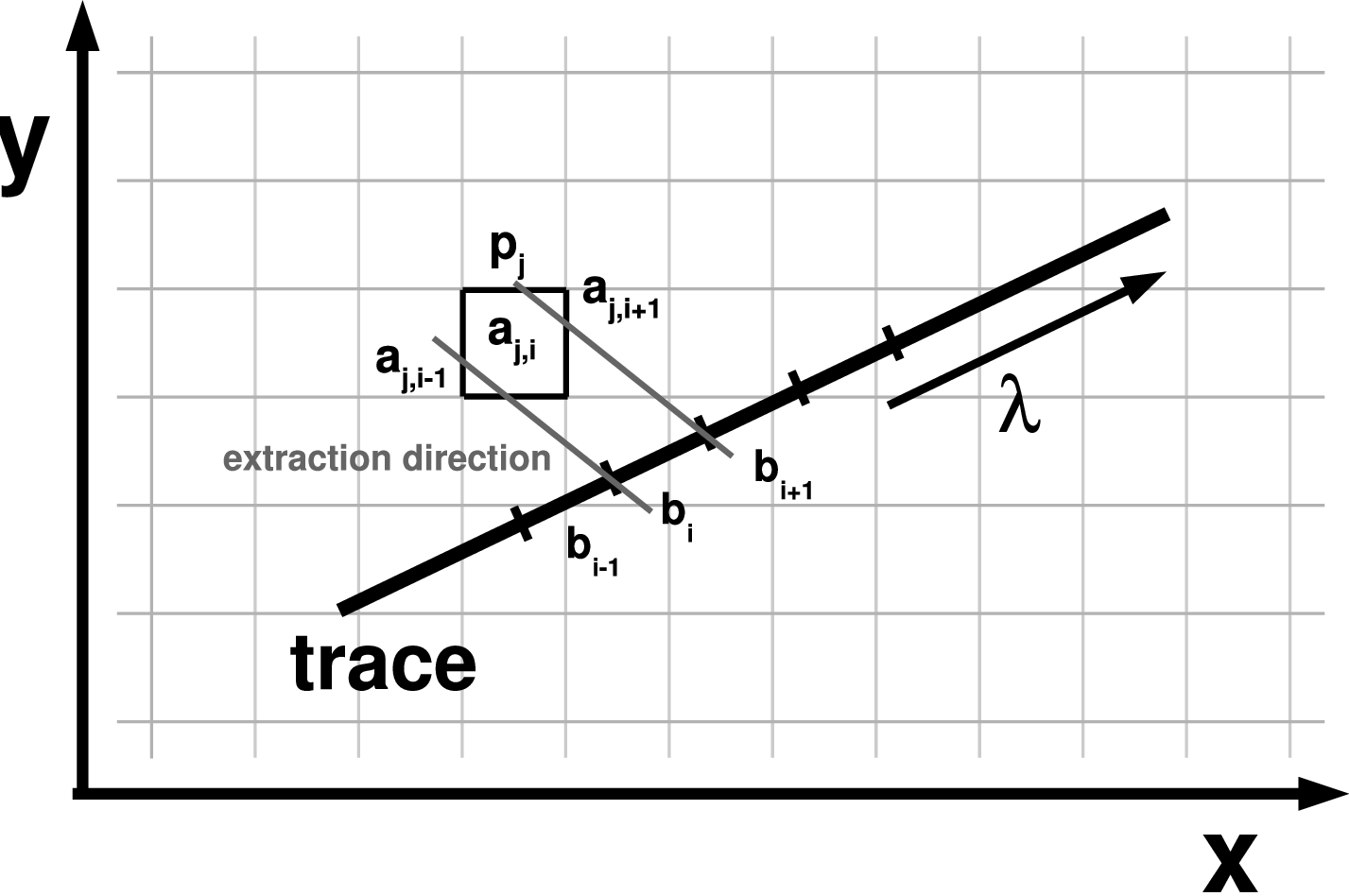}
\caption{The projection of pixel $p_j$ onto the set of spectral bins
$[\ldots, b_{i-1}, b_i, b_{i+1},\ldots]$. For the given trace and extraction direction (gray, thin lines), $p_j$
contributes with the weight ($=area$) $a_{j,i-1}, a_{j,i}, a_{j,i+1}$ to the bins
$b_{i-1}, b_i, b_{i+1}$.
\label{fig:extraction}}
\end{figure}
The spectral extraction in aXe aims at avoiding multiple rebinning of the pixels
and processes the data from detector to the 1D spectra in a direct manner.
This is achieved by tying the spectral information to the pixels. 
In detail, the aXe extraction of a single object spectrum from a slitless image involves 
the following steps:
\begin{enumerate}
\item on the basis of the reference position, the local trace and dispersion solutions are determined;
\item the extraction geometry (the direction of constant wavelength and slit width, see Sect.\ \ref{sect:extr_geom})
and the extraction area are determined;
\label{ext_one}
\item for each pixel $p_j$ in the extraction area,
the distance to the trace in the cross-dispersion direction,
the distance along the trace and the effective wavelength associated to the pixel
are computed;
\item each pixel $p_j$ is flat-fielded according to its associated wavelength (see Sect.\ \ref{sect:flatfield});
\item a set of spectral bins $[b_0, b_1, b_2, \ldots, b_n]$ is created.
Each bin $b_i$ covers the wavelength range
\begin{eqnarray}
\Delta \lambda = [(\lambda(i) - \lambda(i-1))/2, (\lambda(i+1) - \lambda(i))/2]
\label{eqn:lambda}
\end{eqnarray}
where $\lambda(i)$ is the grism or prism dispersion solutions given in
Eqns.\ \ref{eq_grdisp} and \ref{eq_prdisp}. The sequence of integer numbers $i$ is chosen
such that the entire sensitive wavelength range of the instrument is covered;

\item the area of each pixel $p_j$ is projected
onto the set of spectral bins $b_i$ in order to determine the fractional contributions
$a_{j,i}$. As can be seen in Figure \ref{fig:extraction}, the quantities $a_{j,i}$
are usually $zero$ for all except a few bins. The extraction or cross-dispersion
direction is optimized for each object (see Sect.\ \ref{sect:extr_geom}) and is, as
shown in Fig. \ref{fig:extraction}, not necessarily perpendicular to the trace;
\item the spectrum value of the spectral bins with {\it non-zero} contributions
$a_{j,i}$ are re-calculated using weighted summation
\begin{eqnarray}
w'_{b_i}& = & a_{j,i} + w_{b_i}\label{eqn:weight}\\
s'_{b_i}& = & \frac{a_{j,i} * s_{p_j} + w_{b_i} * s_{b_i}}{w'_{b_i}}
\label{eqn:wsum}
\end{eqnarray}
where $w'_{b_i}$, $s'_{b_i}$ and $w_{b_i}$, $s_{b_i}$ are the new and old values for the
spectrum value and weight for bin $i$, respectively and $s_{p_j}$ is the
spectral flux in pixel $p_j$.
\label{ext_seven}
\end{enumerate}

After the extraction the spectral bins for each object, ordered in wavelength,
are written to a FITS table \citep{po94}. These resulting spectra contain for each spectral bin
the wavelength $\lambda$, the extracted spectrum in $[electrons/s]$, 
the error in $[electrons/s]$, the flux in $f_\lambda$, the flux error and the total weight.

The wavelength values, although derived from spectral bins $b_i$ which have associated wavelength ranges as
described above, are {\it not} just assigned the mean value of the spectral bins $\lambda(i)$
(see Eqn.\ \ref{eqn:lambda}), but are computed using weighted summation
as in Eqns.\ \ref{eqn:weight} and \ref{eqn:wsum} (with $\lambda_{p_j}$, the effective
wavelength associated to pixel $j$, instead of $s_{p_j}$).

The error for each spectral bin is computed by propagating the initial error for each detector
pixel through all aXe reduction
steps using basic error propagation. If not explicitly given in an image extension of the slitless image,
aXe derives an error image on the fly, using a typical detector noise model (readout noise and photon shot
noise).

The total weight of a spectral bin is the co-added area of its pixel contributions.
Since aXe does not use pixels which are marked as defective, the weights for a spectral
bin can vary. For HST images, such bad pixels (e.g. dead pixels or pixels affected by cosmic rays)
are indicated by the value of the data quality image extension.

\subsection{Extraction geometry}
\label{sect:extr_geom}
\begin{figure}[t]
\includegraphics[angle=0, width=0.45\textwidth]{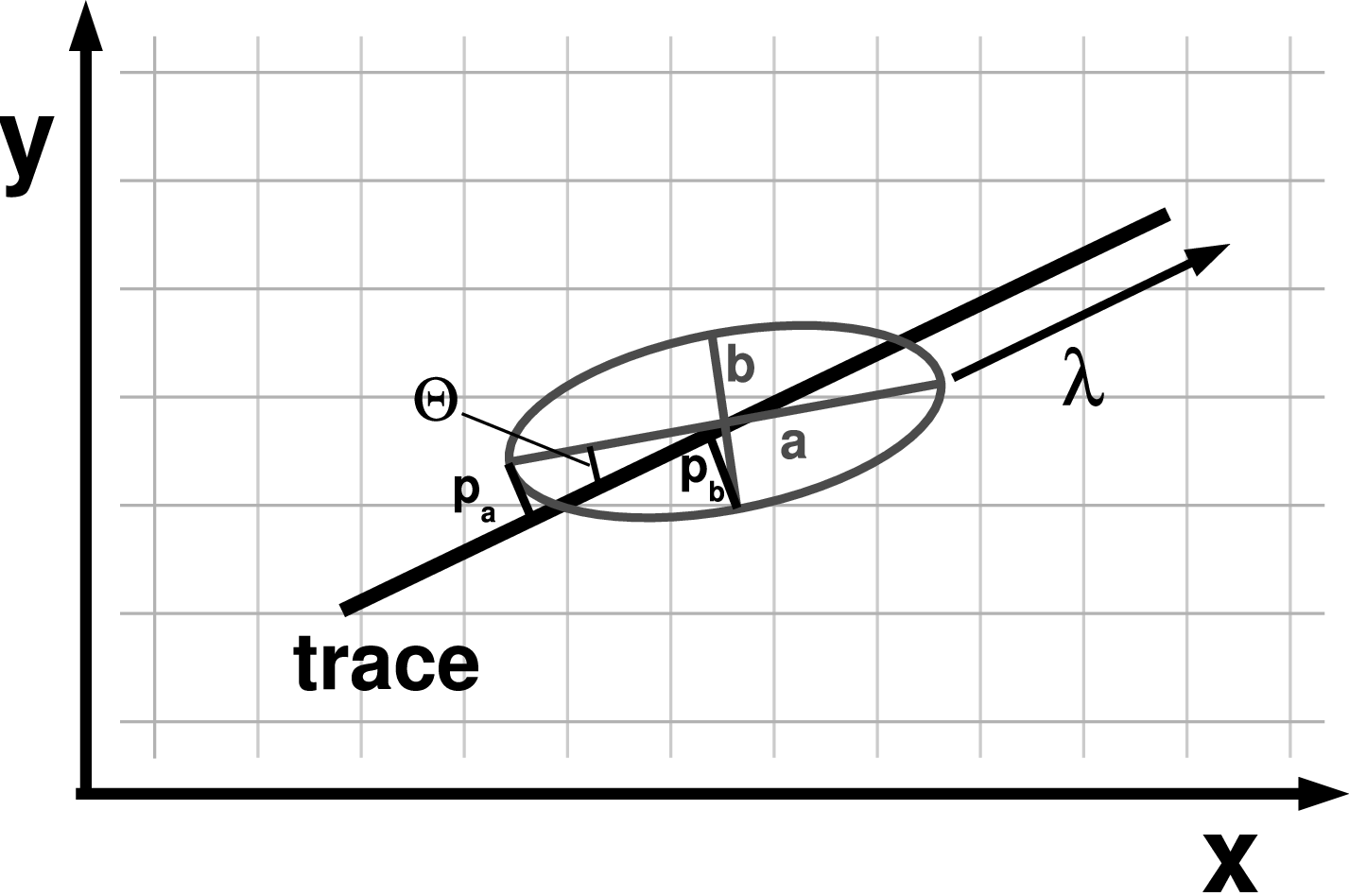}
\caption{An illustration of the parameters for defining the extraction geometry. The object morphology is
described
as a 2D Gaussian with major and minor half axes $a$ and $b$ and the angle $\Theta$, here defined with respect
to the dispersion or trace direction. The quantities $p_a$ and $p_b$ are the projection of the major and
minor half axis onto a plane perpendicular to the trace.\label{fig:extr_geom}}
\end{figure}
Every object which produces a slitless spectrum acts as its own
{\em virtual slit} (see Appendix in Freudling et al.\ 2008 for a detailed discussion), which in general is not
aligned with the detector grid
or the spatial direction of the disperser. The shape of this virtual
slit must be taken into account if a 1D spectrum is to be extracted
from the slitless image or  a rectified 2D spectrum (analogous
to a long slit spectrum, see Sect.\ \ref{sect:axedrizzle}) is to be
generated. Since there is no slit or mask which defines or limits the extraction geometry,
the 'slit orientation' and the 'slit length' translate in slitless spectroscopy to the extraction
direction (the direction of equal wavelength) and the extraction width (the number of pixels
on both sides of the trace which are co-added), respectively. These quantities must be set for
each object during the spectral extraction.

aXe offers several methods to define the extraction geometry and thus to extract 1D spectra
from a slitless image. Most of the methods link the extraction geometry to the object
shape in order to optimize the extraction.
Figure \ref{fig:extr_geom} illustrates the quantities involved in
defining the extraction geometry for an arbitrary object. In a very basic model
an object is described as a 2D Gaussian with major axis size $a$, minor axis size $b$
and the orientation of the major axis $\theta$, which is defined with respect
to the trace here. The projections of $a$ and $b$ onto the plane perpendicular to
the dispersion or trace direction are named $p_a$ and $p_b$, respectively.
For this object shape, aXe offers spectral extraction with:
\begin{enumerate}
\item a fixed extraction width $w=const.$ given by the user and an extraction direction perpendicular to the 
dispersion direction;
\label{seq:case_one}
\item a variable extraction width scaled as $w=k*MAX(p_a, p_b)$ ($k$ given by the user) and an
extraction direction perpendicular to the dispersion direction;
\item a variable extraction width scaled as $w=k*a$ ($k$ given by the user) and an
extraction direction parallel to $a$;
\label{seq:case_three}
\item as in \ref{seq:case_three}, but switching to $w=k*b$ and an extraction direction
parallel to $b$ if $p_b> p_a$.
\label{seq:case_four}
\end{enumerate}

Depending on the science case the user may prefer a specific extraction method, e.g (\ref{seq:case_one})
if the primary targets are point-like or (\ref{seq:case_four}) in a typical survey
situation with galaxies of differing sizes and random orientation.
 
\subsection{Flat-fielding slitless spectroscopic data}
\label{sect:flatfield}
\begin{figure}[t]
\includegraphics[angle=0, width=0.5\textwidth, totalheight=0.5\textwidth]{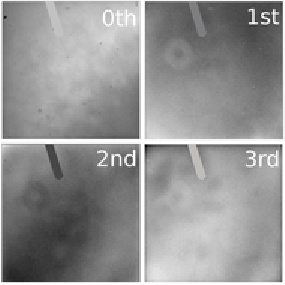}
\caption{The four coefficient images of the HRC/G800L flat-field cube.
In the region (top left) obscured by the coronographic finger the coefficients
result in a flat-field value $1.0$ for all wavelengths.
\label{fig:HRCff}}
\end{figure}
Flat-fielding is an essential data reduction step in creating scientific
quality images of uniform sensitivity. In direct imaging, flat-fields
for each filter are constructed using
homogeneously illuminated images, and a single flat-field correction
yields equal sensitivity. These direct imaging flat-fields are valid
for radiation in the restricted passband defined by the combination of optical
elements. A slitless spectrum, however, can occur anywhere on the detector, and the
flat-field correction has to be performed
for each pixel as a function of incident wavelength; this leads to the
use of a flat-field cube whereby the wavelength behaviour of the 
flat-field is described for each pixel by the 3$^{rd}$ dimension in addition
to the two spatial dimensions of the detector. 
To characterize  the wavelength dependent behaviour, a series of direct image
flat-fields taken at various central wavelengths is used.

For ACS, we decided to combine the wavelength information in all direct image
flat-fields by fitting a polynomial to the series of  wavelength dependent
flat-field values for each pixels. The flat-field cube then contains the
polynomial coefficients in the third dimension. This method is applicable
for the ACS slitless modes since the variation of the flat-field as
a function of wavelength is smooth \citep{bo02}.

There are several reasons why the flat-field cube derived from direct image
flats only does not provide a perfect correction.
The large scale flat-field characteristic of the grisms and prisms is likely
somewhat different from that of the direct image flats used to generate the
flat-field data cube. Moreover the direct image flat-fields produce a flat
image for a homogeneously illuminated image of the sky, but the plate scales
and hence the effective pixel size vary significantly from one corner of the
detector to another \citep{co02}.

In order to correct for these additional, field dependent effects it is necessary to
apply an empirical large scale correction to the flat-field cube, which
is determined from observations of a flux standard
star at various detector positions. The final flat-field cube allows the application of
a single sensitivity curve for the conversion to absolute flux units which
is independent of the object positions.

A detailed description on flat-fielding in slitless spectroscopy and a
discussion on the use of the available direct image flats is given in
\cite{wa05}. Figure \ref{fig:HRCff} shows a representation of the ACS HRC/G800L
flat-field cube. Its sub-panels show the four layers with the polynomial
coefficients as indicated in the labels. The flat-field value for the pixel
$(i,j)$, which is exposed to light of wavelength $\lambda $ is computed as
\begin{eqnarray}
FF(i,j,x) &=& a_0(i,j) + a_1(i,j) x  +\\ \nonumber
          & & a_2(i,j) x^2  + a_3(i,j) x^3
\label{eqn:ff}
\end{eqnarray}
where $a_k(i,j)$ is the coefficient from the $k^{th}$ layer of the flat-field cube,
$x$ the normalized wavelength,
\begin{eqnarray}
x = (\lambda - \lambda_{MIN})/(\lambda_{MIN}-\lambda_{MAX})
\end{eqnarray}
and $\lambda_{MIN}$ and $\lambda_{MAX}$ are the minimum and maximum wavelengths of the direct
image flat-fields used in constructing the slitless flat-field cube, respectively.

\begin{figure}[t]
\includegraphics[angle=-90, width=0.5\textwidth]{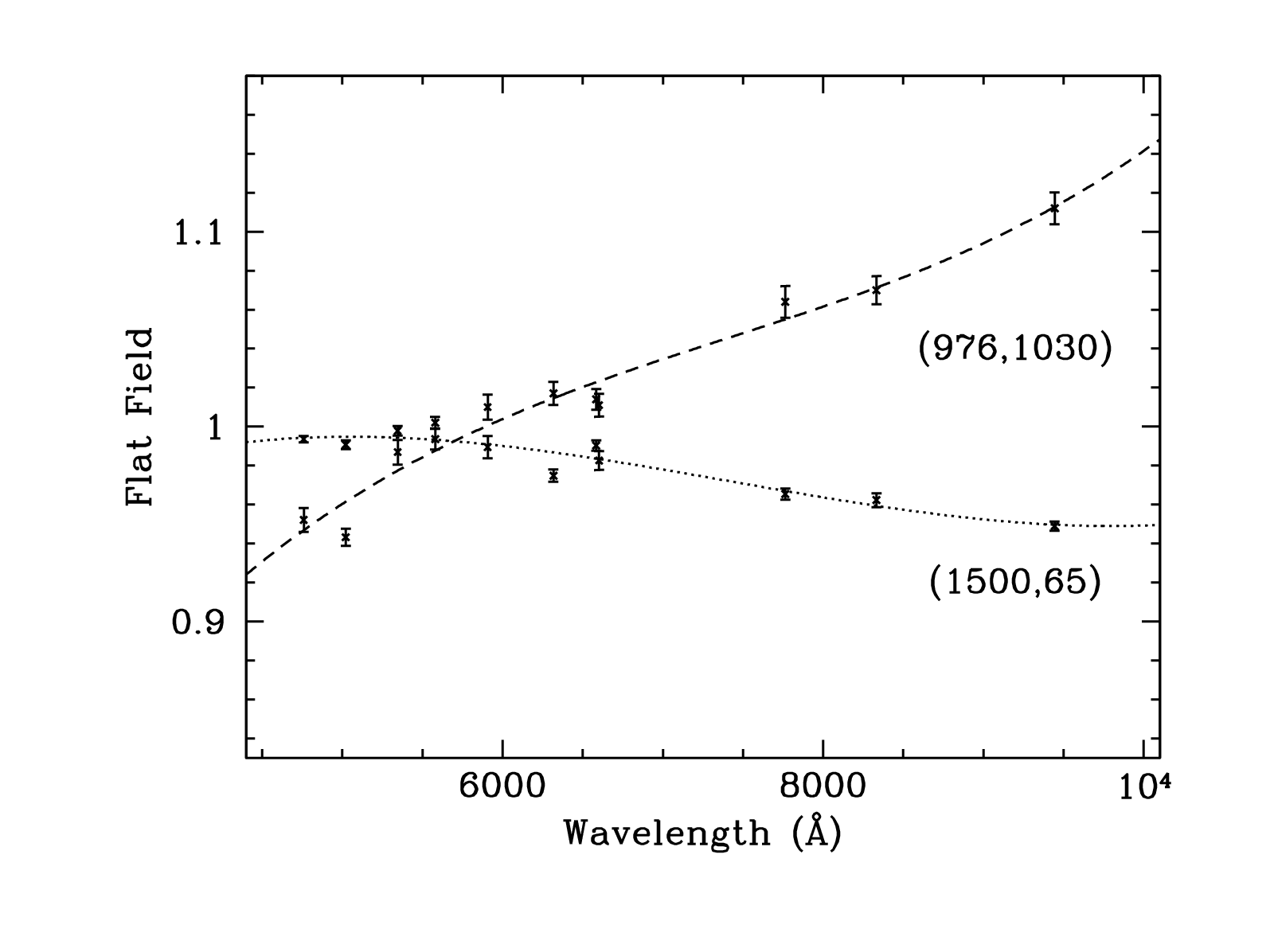}
\caption{The flat-field values as a function of wavelength for two ACS/WFC pixels
(CCD \#1) derived from direct
image flat-fields. The coefficients of the polynomial
fits (shown as lines) to the data are stored in the flat-field cube.
\label{fig:FFpixels}}
\end{figure}

Figure \ref{fig:FFpixels} illustrates the flat-field as a function
of $\lambda$ for
two selected pixels on the ACS/WFC CCD \#1. The symbols show the
values from the direct image flats taken through the ACS filters
for a $2\times2$ binned region, with the rms on the mean as errors.
The lines show the polynomial fits to the the data and thus the
wavelength dependent flat-field correction which is applied
to the data values measured in these two pixels.

\subsection{Background determination}
\label{sect:background}
aXe has two different methods to handle the background emission from
the sky. The first method incorporates a {\it global} background subtraction
before starting the extraction process, the second uses {\it local}
sky background estimates. Further enhancements to the basic methods
described here, e.g. applying a 2D correction to the master sky when
the background shape differs \citep{ma08}, are possible.

\subsubsection{Global background}
\label{sect:glob_backgr}
The method of removing a global sky background uses a 2D master sky image
which is automatically scaled to the background level of each individual slitless
image. When determining the scaling factor, pixels on and close to the spectral
traces are masked out in order to prevent an overestimate of the
background level. The spectral extraction is then conducted after
subtracting the scaled master background from the slitless images.
Master sky images can be generated by e.g. combining multiple slitless
images with a rejection algorithm which excludes pixels dominated by
object spectra.

\subsubsection{Local background}
\label{sect:loc_backgr}
The local background estimate uses the background pixel values close to the
spectral traces to determine individual background levels.
For each pixel which is part of an object spectrum, a background
value is determined by interpolating between the user-defined number of adjacent background pixels.
Figure \ref{fig:composit_image} shows the cutout of a local background image
in panel c. Interpolated background pixels are visible at the position of the
spectra, the background pixels used for the interpolation remain unchanged,
and pixels neither part of a spectrum nor used as background appear white.

The local background subtraction can even be applied in addition to
the global background. Such a procedure is useful in slitless modes such as NICMOS
G141, which display temporal variations of the background
(see Freudling et al.\ 2008).

\subsection{Discussion of the extraction method}
\label{sect:extr_discussion}
The aXe approach to extract spectra is different from
the long slit or multi-slit approaches. The method of equipping the pixels with spectral information
(e.g. wavelength, trace distance, see Sect.\ \ref{sect:spec_extr}) and then 'filling' them
into initially empty spectral bins has several advantages. Firstly, the re-use of an
individual pixel in several spectra, which is necessary in regions of spectral overlap,
is straight forward. Several wavelength values $\lambda_0, \lambda_1,\lambda_2, \ldots$ are given
to the pixel $p_i$ when co-added to the spectra of the sources $0, 1, 2, \ldots$.
Secondly, the virtual slits defined by the objects themselves can be flexibly addressed
via the various geometrical methods of attributing wavelength values to the pixels
(Sect.\ \ref{sect:extr_geom}). Hence aXe is well equipped to adjust to the
variety of object morphologies encountered in typical survey scenarios.
Thirdly, the techniques used in the aXe spectral extraction are
geometrical descriptions (e.g. for the trace and wavelength solutions or attributing
wavelength values to the pixels) and weighted summation (when co-adding pixels to the
spectra), which both are rather simple and mathematically well described.

\section{Contamination}
\label{sect:contamination}
In slitless spectroscopy an overlap of spectra occurs very frequently.
This overlap is generally referred to as 'contamination', as it complicates
the task of extracting a clean spectrum of a given single object.
There exist several types of contamination.
Firstly, objects close in the spatial direction can cause contamination.
Such contamination can also occur in spectroscopy with slits or
masks. Secondly, objects that are well-separated on the sky can still contaminate
each other in the dispersion direction. In spectroscopy with masks, such
contamination is avoided by a careful object selection or, equivalently,
mask design. Thirdly, for grism dispersers (e.g. ACS/G800L, NICMOS/G141) many
spectral orders ($0^{th}$, $1^{st}$, $2^{nd}$, $-1^{st}$ \ldots) are typically
visible in the FOV of the detector, and contamination can occur across different
spectral orders at distances of several hundred pixels.

Contamination is impossible to avoid in typical slitless
spectroscopy observations at optical/NIR wavelengths, such as the HST slitless modes
ACS/WFC/G800L, ACS/HRC/G800L, NICMOS/G141. Consequently, the data reduction software
must be able to collect and process contamination information in
order to deliver data that can be interpreted scientifically.

aXe handles contamination in two different ways \citep{kue05b}.
Both use predictions based on the knowledge of the source
positions. The so called {\it geometrical contamination} scheme simply
records which spectral bins are affected by contamination.
The {\it quantitative contamination} scheme estimates for each spectral
bin the contribution from neighbouring sources.

\subsection{Geometrical contamination}
\label{sect:qualcont}
For geometrical contamination the areas that are covered by the various
spectral orders of all objects are marked. Then for every pixel in
each spectrum the information on how many other spectra fall on this pixel
is processed through the extraction. The final result is the number
of other spectra that fall on the pixels which were combined to a spectral
bin. Geometrical contamination is computationally cheap and allows the
selection of regions of spectra which are not occupied and therefore
contaminated by other spectra.

\subsection{Quantitative contamination}
\label{sect:quantcont}
Quantitative contamination aims at computing directly for
every spectral bin the contaminating flux from other sources.
This is done by modelling the dispersed contribution of every object
to the slitless image. Based on this model information the contaminating
flux for each pixel is recorded and processed through the spectral
extraction. The result of quantitative contamination is, for each
extracted spectrum, an associated spectrum of the contaminating flux.

aXe generates a slitless image model re-using the configuration 
and sensitivity files from the extraction, which guarantees a symmetry
between the extraction and the modelling in quantitative contamination.
The morphological and spectral object properties are derived from the
direct images, and aXe has two different methods to employ these properties
in an {\it emission model}, which are discussed in the next Sections.

An even better refinement and better estimates would be expected from an
iterative approach which uses the direct image information initially,
but then the extracted spectra for more refined contamination estimates.
This procedure, which is computationally intensive and does not
necessarily converge, is currently not implemented.

\subsubsection{Gaussian emission model}
\label{sect:gaussmodel}
The Gaussian emission model approximates the object morphology
with 2D Gaussians, whose major and minor axes and position angle
are already available in the spectral extraction  (see Sect.\ \ref{sect:extr_geom}
and Fig.\ \ref{fig:extr_geom}). As spectral energy distribution
this emission model uses one or, if available, several AB-magnitudes at
different wavelengths measured on the associated direct images.
Beyond the wavelength
range covered by the AB-magnitudes the object 'spectra' or spectral energy
distributions are extended as flat in $f_{\lambda}$. Figure \ref{fig:composit_image}d
shows the slitless model generated for quantitative contamination using
the Gaussian emission model.

\subsubsection{Fluxcube emission}
The Fluxcube emission model derives both the object morphologies and
the spectral information directly from the direct images. For this purpose
a so called {\it fluxcube file} is generated for each slitless image. Such a fluxcube
file is a multi-extension FITS image which contains at least one corresponding
direct image cutout, converted to $f_{\lambda}$, and a {\it segmentation image}.
The segmentation image, essential to avoid self-contamination, marks the pixels
contributing to each object. The values of the pixels on which an object in the
associated source catalogue was detected contain the object index.
A segmentation image is easily produced by SExtractor \citep{be96}
(parameter '{\tt CHECKIMAGE\_TYPE SEGMENTATION}'). As for the Gaussian emission
model, the  fluxcube emission model extends the object 'spectra' as flat in $f_{\lambda}$.

\subsection{Contamination example and discussion}
\begin{figure}[t]
\includegraphics[angle=-90, width=0.5\textwidth, totalheight=0.45\textwidth]{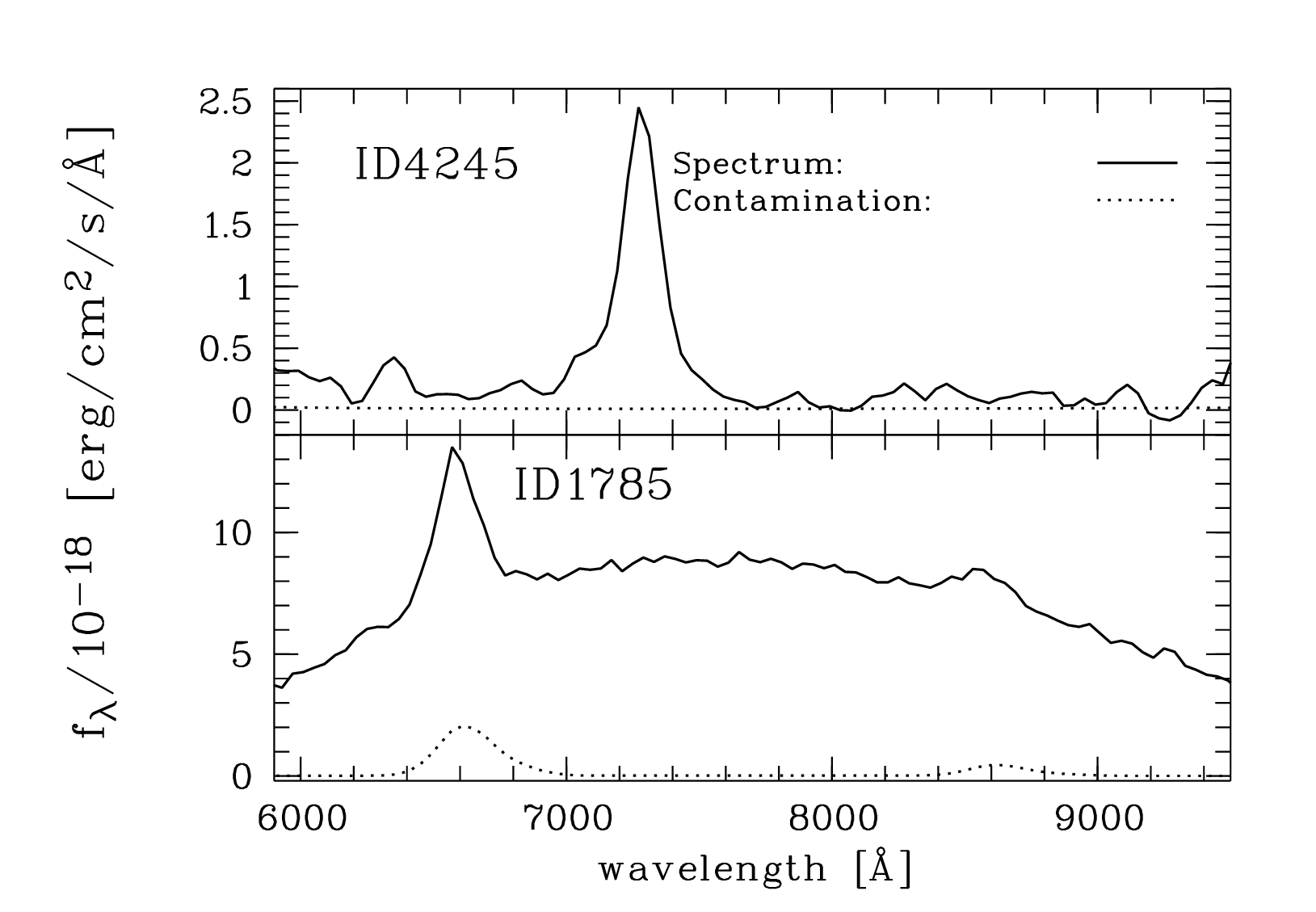}
\caption{Extracted ACS/WFC/G800L spectra (solid lines) and quantitative contamination estimates (dotted lines,
Gaussian emission model) for two spectra extracted with aXe. The contamination 
estimate in the lower spectrum indicates that both emission lines results
from an overlap with the spectrum of other sources.\label{fig1:quantcont}}
\end{figure}
Figure \ref{fig1:quantcont} illustrates how quantitative contamination aids in
the interpretation of extracted spectra. The two panels show source spectra (solid lines) and
quantitative contamination information (dotted lines, Gaussian emission model)
for two objects extracted from HST ACS/WFC/G800L slitless data.
The upper source (named {\sl ID4245} internally) has a strong
emission line at $\sim 7300$\AA,  and the
lower ({\sl ID1785}) two emission features at $\sim 6600$ and $\sim 8600$\AA.
The geometrical contamination scheme raises the contamination flag for all spectral
bins in both spectra. The quantitative
contamination scheme is able to estimate for the upper
spectrum a negligible contamination from other sources and for the lower spectrum
at least a strong contributions from neighbouring objects to both emission features.
The two spectra in Fig.\ \ref{fig1:quantcont} clearly demonstrate the
need for quantitative contamination information. This is especially
essential for deep slitless surveys such as GRAPES \citep{pir04}, where contamination
by $0^{th}$ orders often resembles emission lines (as in {\sl ID1785})
and a reliable distinction
is essential.

The basic data necessary to extract slitless spectra with aXe (slitless images plus
associated direct images) already allows the usage of quantitative contamination
with little extra effort. A limitation of quantitative contamination is the emission
model, and particularly its spectral part. The crude approximation of the object
spectra with often only a single data point in flux space can still lead to large differences
between the contamination estimate and the actual contamination in the spectra.
E.g. an examination of the spectra {\sl ID1785} in Fig.\ \ref{fig1:quantcont}
reveals that the contamination estimate is $\sim 2$ to $3$ times lower 
than the excess in flux at $\sim 6600$\AA. A better wavelength
coverage with more direct images yields better contamination estimates.

\section{2D-resampling of spectra}
\label{sect:axedrizzle}
As in direct imaging, a typical slitless spectroscopic dataset taken with HST
usually consists of several images of the same region on the sky with only small
shifts or dithers \citep{koe02} between them. The basic aXe extraction
discussed in Sect.\ \ref{sect:axe_strat} is capable of reducing all objects
from every slitless image. However the data for each object must be co-added to
obtain deep spectra with high signal-to-noise ratio.

\subsection{Co-adding individual spectra}
\label{sect:coadd_indiv}
Co-added object spectra can be
derived by combining the 1D spectra extracted from the individual slitless
images. This approach has some disadvantages.
The data is rebinned twice, once for the spectral extraction and
once for combining the spectra. The procedure requires a complex weighting scheme
to take into account masked pixels (e.g. cosmic ray hits) and different exposure
times for the spectral bins. Information in the spatial direction, e.g. the exact
location of an emission feature within an extended source, is lost when combining 1D
spectra. Problem detection and error tracking is more difficult in 1D than in 2D.

\subsection{Co-adding spectra with aXedrizzle}
\label{sect:axedrz_coadd}
With {\it aXedrizzle} (K\"ummel et al.\ 2004; K\"ummel et al.\ 2005) 
we have developed and implemented in aXe a method of combining the individual,
2D spectra into a deep 2D slitless spectrum, which allows subsequent 1D extraction from the deep image. For
2D-resampling and co-addition of the individual 2D spectra we use the drizzle
software \citep{fru02}, which is a standard method of combining HST direct images.
With aXedrizzle the re-gridding to a uniform wavelength scale and a
cross-dispersion direction orthogonal to the dispersion direction is achieved in
a single step. The weighting of different exposure times per pixel and cosmic-ray
affected pixels are correctly handled through the drizzle weights.
The combined 2D spectra can be visually inspected for problem detection.
They reveal fainter features in the spatial direction
not detectable in single images (see e.g Rhoads et al.\ 2004) and can be used for finding
emission line regions in larger, extended objects.

\subsection{The aXedrizzle technique}
\label{sect:axedrz_tech}
\begin{figure}[t]
\includegraphics[angle=0,scale=0.45]{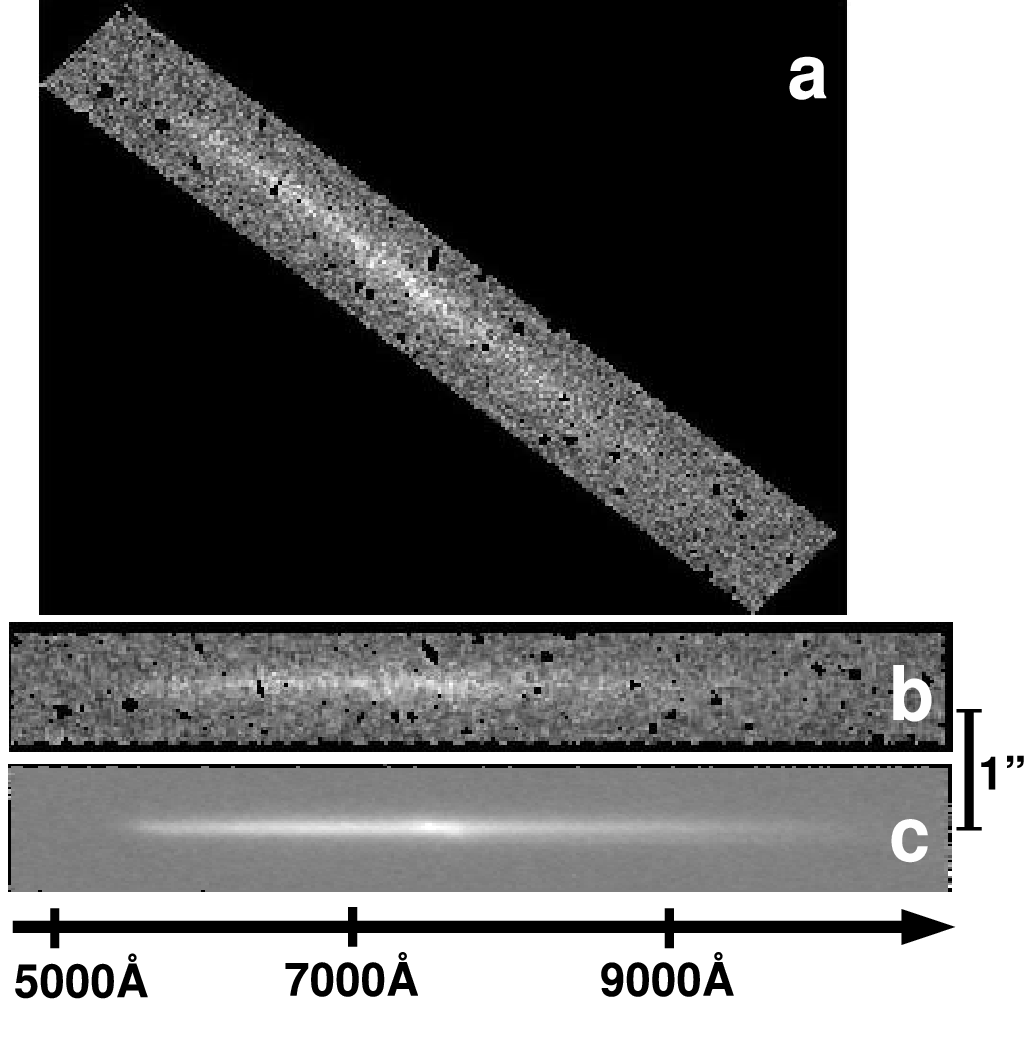}
\caption{An overview of aXedrizzle steps and products. The flat-fielded object pixels
are re-assembled to
a cutout image (a). Drizzling this image leads to an individual 2D spectrum with zero
trace angle, linear dispersion and perpendicular extraction direction (b).
The individual 2D spectra are co-added to a deep, 2D slitless
image (c).\label{fig1:aXedrizzle}}
\end{figure}
As an example, Figure \ref{fig1:aXedrizzle} illustrates the aXedrizzle procedure for ACS/HRC
G800L data from the Hubble Ultra Deep Field
(HUDF; proposal ID 9803, P.I: Beckwith; \\\anchor{http://www.stsci.edu/hst/udf}{http://www.stsci.edu/hst/udf}) 
Parallels observations for one object.
The upper panel in Fig. \ref{fig1:aXedrizzle} shows the cutout image for one object
in one grism image, assembled from the flat-fielded pixels. In the $112$ slitless
images of the entire dataset, the object is located at different positions, which
leads to small variations in trace angle, dispersion solution and image distortions.
For each of the $112$ observations of this object individual drizzle coefficients
are determined which, via the drizzle process, lead to individual 2D grism images,
such as Fig.\ \ref{fig1:aXedrizzle}b, with zero trace angle, a linear, pre-defined
dispersion solution and an extraction direction perpendicular to the trace.
aXedrizzle allows the user to select a specific interpolation kernel for the drizzle routine
in the configuration file. The final 1D spectrum is extracted from the co-added
2D slitless stamp
image Fig.\ \ref{fig1:aXedrizzle}c, which has a combined exposure time of $124ksec$.

\begin{figure}[t]
\includegraphics[angle=0, width=0.5\textwidth]{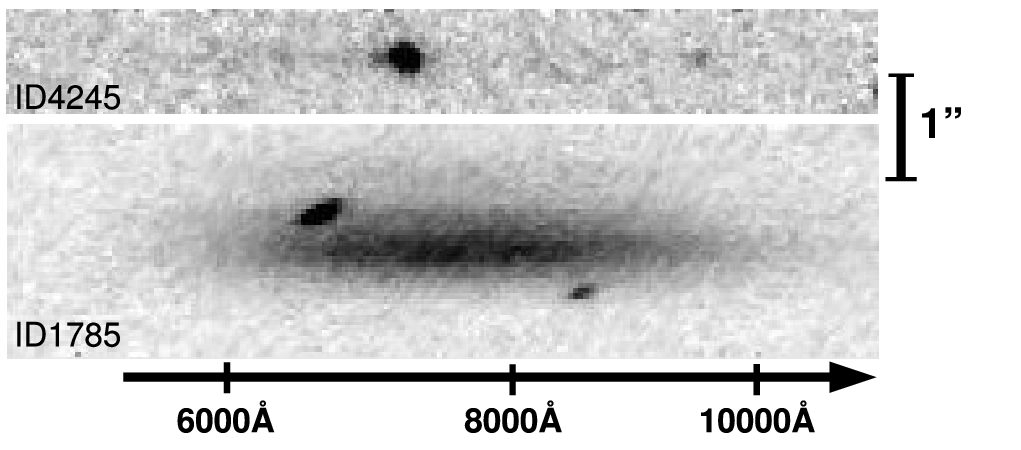}
\caption{The aXedrizzled, deep 2D slitless images for the sources shown in
Fig.\ \ref{fig1:quantcont}. For $ID1785$ (lower panel) the contaminating $0^{th}$
order spectra from other sources are clearly visible.\label{fig1:gristamp}}
\end{figure}
To illustrate the advantages of aXedrizzle, Figure \ref{fig1:gristamp}
shows the aXedrizzle combined, deep 2D slitless images corresponding to the 
spectra in Fig.\ \ref{fig1:quantcont}. The re-sampled image
confirms the intrinsic origin of the emission line feature for the source {\sl ID4245} in
the upper panel and even indicates the presence of additional emission lines. In the
lower panel, it is obvious that the object {\sl ID1785} itself has a smooth spectrum.
Both 'emission features' arise solely from two bright, offset spots, which are
identified as contaminating zeroth order spectra from two different objects.

Besides the main science information, aXedrizzle also processes other data such as
the errors, contamination (see Figs.\ \ref{fig1:quantcont} and \ref{fig1:gristamp})
and weights (for optimal weighting, see Sect.\ \ref{optimal_weighting}). All data
is stored in the layers of the deep 2D slitless stamp, which is a multi-extension
FITS image, and used in the 1D extraction. As in the original drizzle
method, the weight image is composed as one image extension which reflects the
total exposure time for each pixel.

The 2D resampling and co-adding in aXedrizzle introduces correlated
errors between neighboring pixels. However in \cite{fr08} it was shown that
the independently processed noise from aXedrizzle is in good agreement with
empirical noise estimates derived from DER\_SNR \citep{st08}. This confirms
not only the accuracy of the errors from aXedrizzle but also justifies
their further use, e.g. in optimal weighting (Sect.\ \ref{optimal_weighting}).

Local background estimates are processed by aXedrizzle in the same way as the 
object data, and for each deep 2D slitless stamp a 'deep' 2D background stamp image
is generated from the local background estimates on the individual images. As in the
processing of individual slitless images (Sect.\ \ref{sect:background}), the background
subtraction is done during the 1D extraction of the spectra.

Only the 2D combining of the first order spectra is
currently implemented in aXedrizzle. In all grism modes supported by aXe, the sensitivity
contrast between the first and all higher orders is very large, and
the focus on the first order does not imply an important loss of scientific
content which the slitless data contains.

\subsection{Data with different roll angles}
\label{sect:axedrz_roll}
Apart from dithering, HST slitless data can also be observed with different roll angles,
which translates to a different orientation of the traces on the sky. Certainly the simple
method of co-adding 1D spectra (Sect.\ \ref{sect:coadd_indiv}), but also aXedrizzle, is
capable of combining such data. As well as the spectra, the individual contamination patterns,
which usually change for significant roll angle differences, are combined. In general, this leads
to combined spectra with larger and more complex contamination. Also care must be taken
when combining data of extended objects, since the object size in the spectral direction,
and thus the spectral resolution, varies for significant roll angle differences.
In addition, more complex objects, such as late-type galaxies with HII regions, 
can display very different spectra at different roll angles.
For point sources, however, it is possible to combine the contamination-free parts
of the data only and to derive contamination free spectra.

\section{Optimal weighting in aXe}
\label{optimal_weighting}
The optimal weighting technique, originally introduced by
\cite{ho86} and \cite{ro86}, aims to enhance the signal-to-noise ratio of spectra by
attributing lower weights to pixels which contain only a small fraction of the 
object flux. The basic equation of the spectral
extraction using optimal weights is (see e.g.\ Rodriguez-Pascual et al.\ 1999):
\begin{eqnarray}
f(\lambda) = \frac{\sum_x \left[ f(x, \lambda) - b(x, \lambda)\right]*\frac{p(x,\lambda)}{\sigma(x,\lambda)^2}}{\sum_x \frac{p(x,\lambda)^2}{\sigma(x,\lambda)^2}} \label{opt_noise}
\end{eqnarray}
where $x$ is the coordinate in spatial direction, $\lambda$ the coordinate in the spectral direction,
$f(x, \lambda)$ the data value at pixel $(x,\lambda)$, $b(x, \lambda)$ the background value at pixel $(x,\lambda)$,
$\sigma(x, \lambda)$ the noise value at pixel $(x,\lambda)$, $p(x, \lambda)$ the extraction profile at pixel $(x,\lambda)$
and $f(\lambda)$ the extracted data value at $\lambda$.

In the original descriptions of optimal weighting, the extraction profile
$p(x, \lambda)$ is computed from the object spectrum itself.
For several reasons this approach is not generally applicable
in slitless spectroscopy. The signal-to-noise ratio of the sources is
often too low to determine an individual extraction profile.
The contamination phenomenon does not permit an automatic and reliable 
generation of extraction profiles for all sources. An iterative approach
on the sometimes hundreds or even thousands of spectra on an ACS slitless image
would require too much computing time.
Also averaging the profile in the spectral direction  does
not give representative results for special object classes, e.g. 
sources with emission lines.

In aXe we have developed an optimal weighting which is robust and practical
for all sources. To compute extraction profiles, the optimal
weighting in aXe uses the models for the dispersed
objects, which were introduced in Sect.\ \ref{sect:quantcont} as the basis of
the slitless models and hence the quantitative contamination technique.
The source-specific models computed from the direct image (using either the
Gaussian or the Fluxcube emission model) 
are a good basis for
calculating the quantity
$p(x, \lambda)$ in Eqn.\ \ref{opt_noise}.
 
The 2D models are also used as an input to calculate the pixel
errors $\sigma(x, \lambda)$ according to the typical detector noise model:
\begin{eqnarray}
\sigma(x, \lambda) = \sqrt{ mod(x, \lambda) + b(x, \lambda) + rdnoise^2}
\end{eqnarray}
where $mod(x, \lambda)$ and $rdnoise$ are the 2D model value at pixel
$(x,\lambda)$ and the detector readout noise, respectively.

\section{Treatment of Detector Fringing} 
\label{sect:fringing}
Optical interference in a CCD occurs 
between incident light and that back-reflected from the thin 
multi-layer structure. In a back-illuminated CCD, if the 
detection layer becomes transparent, the contribution of 
the reflected light can become a large fraction of the incident 
power. The result is a modulation of the signal by fringes 
whose amplitude depends on the optical properties of the layers. 
The level of fringing and its behaviour with wavelength is 
dependent on the number of layers of the CCD, their 
thicknesses and refractive indices.

In imaging and 
slit spectroscopy fringe maps are used to correct for 
the detector fringing. In the case of slitless spectroscopy, 
parallel to the case for flat-fielding, each pixel can receive 
light of any wavelength over the passband of the instrument, 
and the fringe map must be replaced by a cube. In practice
maps of the individual layer thicknesses, together with their 
optical properties, are used together with a simple 
geometrical optics model to predict the fringe amplitude for 
all pixels for any wavelength. \cite{ma03} applied this 
technique to the STIS SITe CCD and an analogous procedure 
was developed for the ACS WFC and HRC detectors \citep{wa03}.

From a set of monochromatic flat-fields over the range of
significant fringing, the Fresnel equations can be solved 
in terms of the thickness of the CCD layers, adopting 
tabulated refractive indices for the layer materials 
\citep{wa03}. In order to 
correct for the fringing at a given pixel in a slitless 
spectrum, the wavelength extent of each pixel (given by 
the position of the dispersing object and the dispersion 
solution), the CCD thicknesses for that pixel and 
the respective refractive indices are used to compute 
the observed fringing of that pixel.

The correction of the fringe amplitude affecting a slitless
spectrum has been incorporated into the aXe software
and \cite{kun08a} present the application to ACS slitless spectra.
Besides the CCD structure, an important input for 
the fringe calculation is the form of the throughput function for 
the CCD pixels. The pixel throughput function is defined as
the relative distribution of light as a function of wavelength 
falling onto a given CCD pixel. In general this is the convolution 
of the intrinsic source spectrum with the line spread function 
(LSF) of the instrument. The determination of the throughput 
function and the computation of the fringing amplitudes would
thus require the a priori knowledge of the source spectrum, 
which in general is not available. However for continuum sources 
and the theoretical case of pure emission line objects
(an object with a spectrum that consists only of a single, narrow emission line)
the throughput 
function can be estimated.

For continuum sources the LSF dominates the form of the throughput function
everywhere and the influence of the slowly varying source spectra can be 
neglected. Using a Gaussian with 100\AA\ full width at half maximum height
(FWHM) as LSF (similar to the spectral resolution, see
Kuntschner, K\"ummel \& Walsh 2008a) the ACS WFC requires a maximum fringe
correction of 0.1\%. This is negligible in  comparison with other sources of
error (photon noise, flux calibration errors).

For pure emission line sources the throughput 
function is solely determined by the intrinsic object 
spectrum, thus amplitudes of 12\% as measured in the monochromator data
\citep{wa03} and even higher are possible. While such variations
could be detected, in practice the intrinsic 
line broadening in the spectrum dilutes the fringe signal.
As an example, line flux variations of $<5\%$ were detected among a set
of Wolf-Rayet star observations with ACS WFC/G8000L \citep{kun08a}.
If the emission line source is
extended, then mixing of wavelengths in the resulting
slitless spectrum will even further dilute the detected fringe amplitude.

\section{The aXe software package}
\label{sect_axe_software_package}
The aXe extraction software is available for download.
In this section we describe the operational
system of version 1.6, together with two related software packages which are used to
simulate and visualize slitless data.

\subsection{aXe extraction}
aXe is distributed to be installed and loaded
as a local IRAF/PyRAF package\newline
(\anchor{http://www.stecf.org/software/slitless_software/axe/}{http://www.stecf.org/software/slitless\_software/axe/}).
Moreover it is embedded into
the STSDAS
(\anchor{http://www.stsci.edu}{http://www.stsci.edu})
software package 
which is distributed by the Space Telescope Science Institute
(STScI).
The aXe software package is written in {\tt ANSI C} and {\tt Python}
It uses the third party libraries CFITSIO \citep{pe99},
GSL \citep{go06}
and WCSTOOLS \citep{mi06} and is portable between Unix platforms.

The aXe software is split up into several tasks, which perform all
the necessary steps to extract spectra from slitless images. 
There exist two different classes of aXe tasks.
{\it Low Level Tasks} operate on individual images, and all input and
output refers to a particular image. {\it High Level Tasks} operate on datasets, which usually
consist of several slitless images and their associated direct images taken
in the same area of the sky.

Some preparatory steps are required before spectra can be extracted. These steps
mostly deal with generating input source lists from the direct images in the dataset.
The direct images are usually co-added to enhance the signal-to-noise ratio and to remove
pixel defects and cosmics. For HST data, this step is generally performed 
using the MultiDrizzle software \citep{koe06}. An object detection algorithm, 
e.g. SExtractor \citep{be96}, is run on the co-added direct image.
Besides detecting the objects, SExtractor is also capable to deliver all necessary
object parameters such as the positions, Gaussian shapes (for defining the extraction
geometry) and source brightness (for quantitative contamination).
Using SExtractor is not mandatory, however the source lists for aXe must be
formatted in the same way as a SExtractor object list. The positions of the
sources determined on the co-added direct image are first
projected back into the coordinate system of each associated direct image,
and aXe provides a dedicated task to do this step in a typical scenario of
a direct image co-added with MultiDrizzle and a SExtractor source list.
Then the object positions are projected into the coordinated system
of the slitless images as part of the aXe extraction.

\subsection{Simulations with aXe}
\begin{figure}[t]
\includegraphics[angle=0, width=0.37\textwidth]{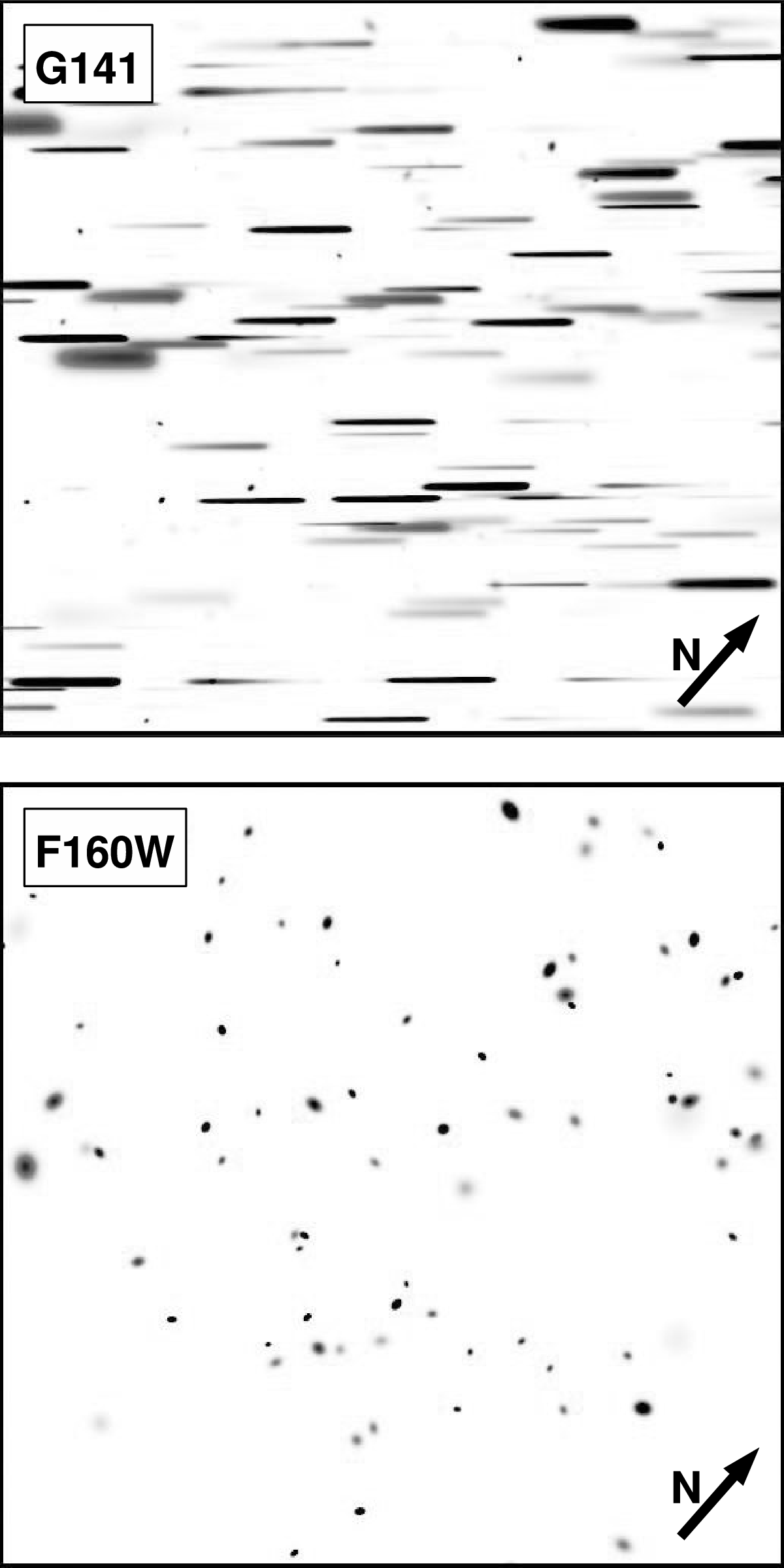}
\caption{aXeSIM simulations of bright objects ($m_{F160W} < 23.0$)
in the NICMOS HUDF field for the HST/WFC3 G141 grism and F160W filter in the upper and
lower panel, respectively. The WFC3/IR FOV is $123''\times137''$.\label{fig1:axesim}}
\end{figure}
The quantitative contamination method introduced in Section \ref{sect:quantcont}
is based on modelling the slitless spectra on input images, either as 2D Gaussians
(see Fig.\ \ref{fig:composit_image}d) or object images. Thus, expanding this modeling
capability to a full simulation 
tool was a natural development. Rather than adding simulations to aXe,
an additional PyRAF package called aXeSIM\newline
(\anchor{http://www.stecf.org/software/slitless_software/axesim/}{http://www.stecf.org/software/slitless\_software/axesim/};
see K\"ummel, Kuntschner \& Walsh 2007), which uses a number of the capabilities of aXe already 
described, was built as a tool for simulating slitless spectroscopic images 
plus their associated direct images.

aXeSIM was designed both to assist users planning slitless observations
and to aid in the analysis of slitless data, such as for determining 
detection limits by adding and extracting model spectra on real images.
For HST, aXeSIM is a valuable tool in the Phase I and II proposal preparation
process, and its usage has already started in Cycle 17. 
A 2D impression of the layout of a target field is
provided, which allows assessment of potential problems arising from 
crowding or spectral overlap. The sensitivity and resolution offered 
by the various slitless modes can be explored in 2D, such as for a 
variety of object morphologies. The optimal pointing and roll 
angle can be chosen for a specific target field. The characterization 
of the instrument used by aXeSIM through a configuration file is 
identical to that used by the extraction package aXe, thus closing
the loop between the simulation and subsequent extraction. 

In the most basic form an object is simulated in the Gaussian emission model
(see Section \ref{sect:gaussmodel}) by a Gaussian shape and one AB-magnitude, transformed and
extended as a spectrum flat in $f_\lambda$. For simulating more realistic objects,
the user can provide 2D image templates for ‘real’ object shapes, build more 
complex spectral energy distributions by specifying magnitude
values at different wavelengths or provide high resolution spectra
as templates, which are shifted in redshift and scaled in flux to 
user-provided values. The basic input for every object is collected in 
a SExtractor-like text table.

As a typical application of aXeSIM, 
Figure \ref{fig1:axesim} shows a noise-free simulation of the
bright ($m_{F160W} < 23.0$) objects in the NICMOS Hubble Ultra Deep 
Field (Thompson et al.\ 2005) for the HST/WFC3 G141 grism and F160W filter 
in the upper and lower panels, respectively. The spectral energy distribution
of the objects is defined via their AB-magnitudes in the filters ACS/WFC/F850LP,
NICMOS3/F110W and NICMOS3/F160W, which all were provided in the original source
catalogue.

\subsection{aXe2web visualization}

\begin{figure}[t]
\includegraphics[angle=0, width=0.5\textwidth]{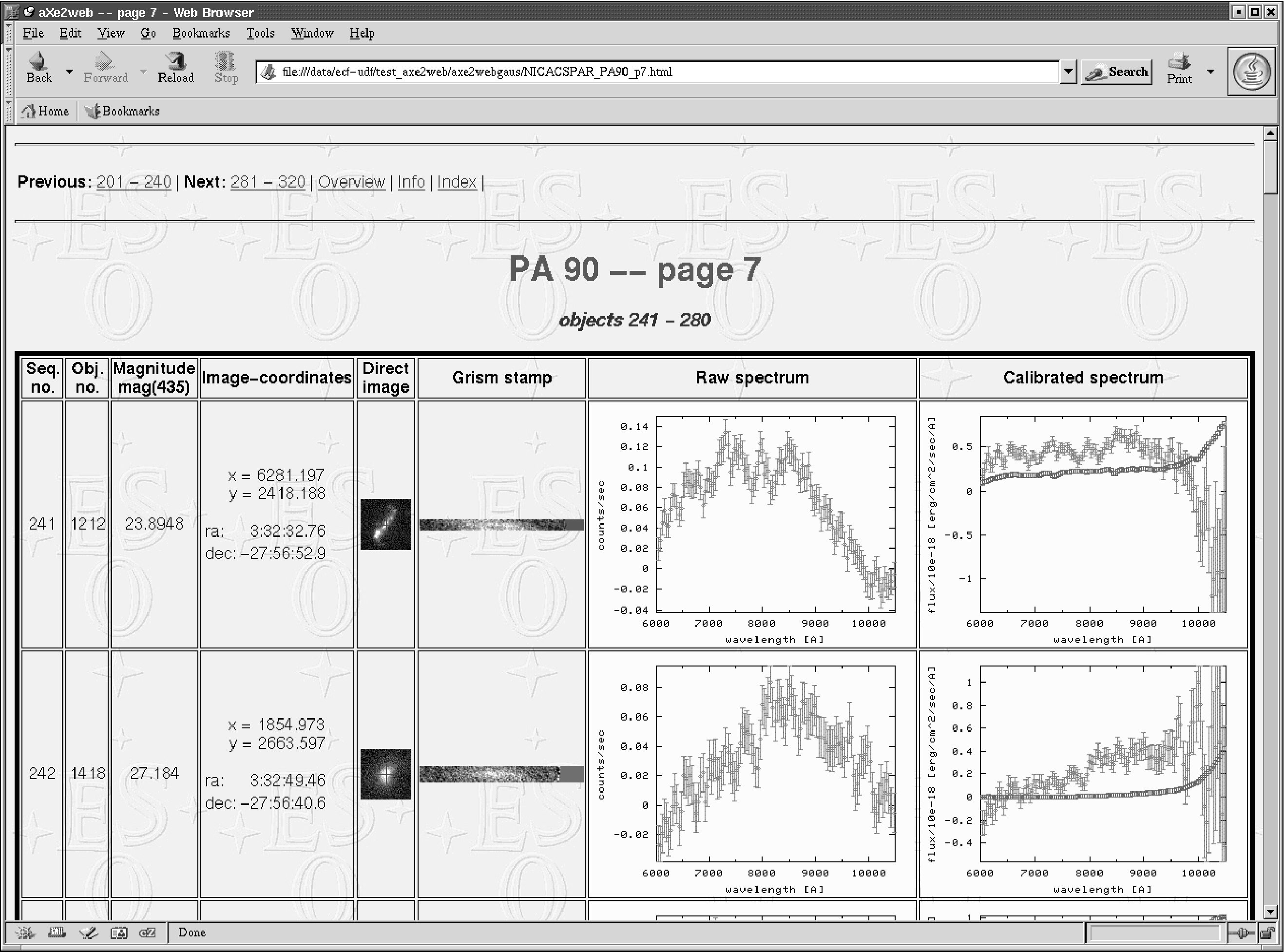}
\caption{An example the visualization of aXe products with aXe2web.
The columns of the
html-table show some basic object information (sequence number, ID number,
brightness, position) followed by a direct image cutout and a 2D slitless image.
The extracted spectra are plotted in $[electron/s]$ and in physical units
($f_\lambda$), overplotted with the contamination estimate.
\label{fig1:axe2web}}
\end{figure}
A deep slitless image (e.g. from ACS/WFC) can contain detectable spectra of
hundreds to over a thousand objects. To visually
inspect them in a fast manner we developed
aXe2web
(\anchor{http://www.stecf.org/software/slitless_software/axe/axe2web.php}{http://www.stecf.org/software/slitless\_software/axe/axe2web.php}),
a tool which produces browsable web pages for fast and discerning examination of
many hundreds of spectra \citep{wa04}.

This additional task to the aXe package takes the aXe
output files and generates linked web pages in html format  
for all sources. Each object produces a line in a table which lists the
reference number, magnitude in the direct image filter, the
X and Y position of the direct object, its Right Ascension and Declination, a
cut-out image showing the direct object, the spectrum stamp image showing the
2D spectrum, a 1D extracted 
spectrum in counts and the same in flux units,
overplotted with the 'spectrum' of the contaminating sources.
Figure \ref{fig1:axe2web} shows an example of two objects extracted
from an ACS/WFC/G800L grism dataset as presented by aXe2web. 

\section{Conclusion and outlook}
aXe is not the only software for extracting HST slitless spectra.
For ACS/WFC, an independent method has been described
in \cite{dr05}. {\tt MULTISPEC} \citep{app05} and the extraction method
described in \cite{ch99} concentrate on STIS slitless spectroscopy.
Quickly after the first release in 2002, aXe
developed into the standard reduction package for all ACS slitless modes.
The methods implemented in aXe specifically for slitless spectroscopy
(e.g. flat-fielding, 2D-resampling with aXedrizzle)
have proven to produce science-ready and well calibrated data for a number of large
programmes such as GRAPES \citep{pir04}, PEARS \citep{ma08} and searches for
high-z supernovae (Riess et al. 2004; Strolger et al. 2004).

By generating instrument-specific configuration and calibration files, aXe 
can also be applied to slitless spectroscopic data from other HST instruments (e.g. NICMOS)
and even from ground based instruments such as
the Wide Field Imager (WFI) mounted on the ESO 2.2-m-telescope. With small 
modifications, aXe has even been used to successfully reduce data taken with 
FORS2 MXU masks (see Kuntschner et al.\ 2005 and K\"ummel et al.\ 2006). aXeSIM is
being used to simulate data for future space missions such as SPACE
(Cimatti et al.\ 2008; now renamed to EUCLID) and SNAP
(\anchor{http://snap.lbl.gov/}{http://snap.lbl.gov/})
that plan to use slitless spectroscopy.

The ability within aXe to automatically extract spectra from slitless data 
was important in the NICMOS Hubble Legacy Archive (HLA) project \citep{fr08}
and aXe served as the central component of the completely automatic Pipeline for 
Hubble Legacy Archive Grism data (PHLAG) (K\"ummel et al.\ 2007; K\"ummel et al.\ 2008).
Due to the modular setup of aXe, additional tasks which handle some 
features specific to NICMOS could comfortably be added in order to 
achieve optimal results leading to an archive of several thousand 
near-infrared spectra
(\anchor{http://www.stecf.org/archive/hla/}{http://www.stecf.org/archive/hla/}).

Besides instrument-specific adjustments also the core of the aXe extraction
software is continually being improved and further developed, with releases
of new versions at an approximately annual rate. New features
and developments that are currently being implemented for aXe (versions 1.7 and 1.8)
and aXeSIM (version 1.2) include 
spectral extraction based on a slit geometry optimized for 
slitless spectroscopy and flux conversion which compensates for the size of 
the direct object (developed in the NICMOS HLA project,
see Freudling et al.\ 2008);
the identification and exclusion of cosmics and hot pixels within aXedrizzle,
similar to MultiDrizzle \citep{koe06} in direct imaging;
the integration of aXeSIM into STSDAS.

The application of aXe to HST data is going to continue with the
installation of the Wide Field Camera 3 (WFC3, Bond et al.\ 2007)
with its three slitless spectroscopic modes (see Tab.\ \ref{spec_modes})
during Servicing Mission 4. As for ACS, aXe is the recommended tool
for reducing WFC3 slitless spectroscopic images. From the ground calibration data
collected during the Thermal Vacuum testing, initial aXe configuration and
calibration files are already available (Kuntschner et al.\ 2008b and
Kuntschner et al.\ 2008c) that were, for example, used in generating the aXeSIM
simulations shown in Fig.\ \ref{fig1:axesim}. Improved calibration will result
from on-orbit data, enabling aXe to extract spectra from all slitless modes of WFC3.

\acknowledgments
Based on observations made with the NASA/ESA Hubble Space Telescope, obtained
from the data archive at the Space Telescope - European Coordinating Facility.\\
We would like to thank S.S.\ Larsen, W. Freudling, and M.\ Demleitner
for their respective contributions to aXe. We are grateful to our anonymous
referee for the numerous suggestions, which helped to significantly improve this paper.




\end{document}